\documentclass{article}
\usepackage[utf8]{inputenc}
\usepackage[english]{babel}
\usepackage{ulem}
\usepackage{tikz}
\usetikzlibrary{positioning, fit, arrows.meta}
\usetikzlibrary{arrows.meta, decorations.pathmorphing, patterns}
\usepackage[square,numbers]{natbib} 
\title{Bibliography management: \texttt{natbib} package}
\usepackage{amsmath}
\usepackage{amssymb}
\usepackage{graphicx}
\usepackage[colorlinks=true, allcolors=blue]{hyperref}
\usepackage{enumitem}
\usepackage{graphicx}
\usepackage{multirow}
\usepackage{float}
\usepackage[letterpaper,top=2cm,bottom=2cm,left=3cm,right=3cm,marginparwidth=1.75cm]{geometry}

\usepackage{amsmath}
\usepackage{graphicx}
\usepackage[colorlinks=true, allcolors=blue]{hyperref}
\numberwithin{equation}{section}
\usepackage{csquotes}

\title{Relativistic Correction to the Magnetic Moment \\ of the Charged Lepton}
\author{Abdelhamid Albaid $^*$}
\date{%
    $^*$Department of Physics, College of Science, University of Ha’il, Ha’il 81451, Saudi Arabia; abdelhamid.albaid@okstate.edu\\%
    \today}

\begin{document}

\maketitle

\begin{abstract}

We derive a general relativistic Hamiltonian valid for both bound and scattering systems by reducing the four-component Dirac equation to a two-component Dirac-Pauli form. Unlike conventional approaches, our formulation includes first-order relativistic corrections in a compact, gauge-consistent expression applicable to arbitrary electromagnetic fields—including non-uniform and time-dependent configurations. As an application, we compute the $\mathcal{O}(\alpha^{2})$ relativistic correction to the Landé $g$-factor in hydrogen-like atoms, revealing a novel $m_{j}^{2}$-dependent term that generalizes the Breit result. This correction is experimentally testable in Penning trap spectroscopy. We further show that relativistic effects become comparable to QFT corrections in highly charged ions where $Z \sim 1/\sqrt{\alpha}$
\end{abstract}

\section{Introduction}
The Dirac equation, central to relativistic quantum mechanics, predicts a magnetic moment of $g_s = 2$ for spin-$\tfrac{1}{2}$ charged leptons. This value arises from decomposing the four-component Dirac spinor into two coupled two-component equations, leading to an interaction energy between the spin magnetic moment and an external magnetic field $\vec{B}$ of the form
\begin{equation}\label{spininteraction}
    H = -\vec{\mu}_s \cdot \vec{B},
\end{equation}
where the spin magnetic moment is defined as
\begin{equation}\label{spin-magnetic}
    \vec{\mu}_s = g_s \mu_B \vec{S}/\hbar,
\end{equation}
with $\mu_B = e\hbar/2mc$ denoting the Bohr magneton and $\vec{S} = \hbar \vec{\sigma}/2$ the spin operator.

The tree-level prediction $g_s = 2$ follows directly from Dirac theory, while higher-order corrections—such as those from quantum electrodynamics (QED)—modify $g_s$. A deviation from $g_s = 2$ is observed for both the electron and the muon, known as the anomalous magnetic moment. J.~Schwinger~\cite{Schwinger1948} first explained this deviation by computing a correction of order $\alpha/2\pi$ arising from virtual photon exchange. Higher-order QED contributions include terms of order $\alpha^2$ with a negative sign~\cite{Sommerfield1957,Petermann1957}. The Standard Model (SM), which incorporates QED, weak, and strong interactions, predicts the muon's anomalous magnetic moment~\cite{Aoyama2020,Aliberti2025}. The updated SM prediction, together with the latest experimental result, indicates agreement between theory and experiment, thereby resolving the long-standing discrepancy~\cite{PhysRevLett131,Aguillard2025}.Here, $\alpha$ is the fine-structure constant,
\begin{equation}
    \alpha = \frac{e^2}{4\pi \epsilon_0 \hbar c} \approx \frac{1}{137}.
\end{equation}

In addition to QED corrections, there are also relativistic corrections to the magnetic moment of the electron obtained by solving the Dirac equation for hydrogen-like atoms, as shown by Breit~\cite{Breit1928}. Standard texts~\cite{Sakurai1967,Kaku1993,Ryder1996} discuss the decomposition and normalization of the large component of the Dirac spinor, thereby elucidating the structure of relativistic corrections.

In this work, we extend that analysis to derive a general Hamiltonian that incorporates first-order relativistic corrections to the magnetic interaction terms. Through systematic normalization of the large two-component spinor, we confirm that all first-order relativistic corrections, including spin and orbital magnetic moments, scale uniformly as \(\mathcal{O}(v^2/c^2)\). Although the magnetic interaction operator (e.g., the Bohr magneton \(\mu_B = e\hbar/2m_ec\)) explicitly contains \(1/c\), this factor is already incorporated into the conventional definition of magnetic moments in cgs units. Consequently, the leading relativistic correction appears consistently at \(\mathcal{O}(v^2/c^2)\). For hydrogen-like systems (\(v/c \sim Z\alpha\)), these corrections equivalently scale as \(\mathcal{O}(Z^2\alpha^2)\), unifying all terms -kinetic, spin and orbital - under the same order.

Our derivation, outlined in detail in Appendices B and C, leads to a Hamiltonian applicable to both scattering and bound-state systems under arbitrary electromagnetic fields. This general applicability of the Hamiltonian to both bound and scattering systems represents a key novelty of the present work.

The Foldy-Wouthuysen (FW) transformation is a unitary method used to diagonalize the Dirac Hamiltonian ($H_D$) in the presence of an external electromagnetic field $A^\mu$. This transformation block-diagonalizes the Hamiltonian through successive unitary rotations, with the upper block corresponding to the effective two-component Dirac-Pauli Hamiltonian $H_{DP}$. The first transformed Hamiltonian is given by
\begin{align}\label{pauli}
H_{\text{DP}} = e^{T} H_D e^{-T} - i e^{T} \partial_t e^{-T},
\end{align}
where $\partial_t = \frac{\partial}{\partial t}$, and $T = -\frac{e}{2m} \vec{\alpha} \cdot \vec{\pi}$ is the generator of the transformation, with $\beta = \gamma^0$, $\boldsymbol{\alpha} = \gamma^0 \boldsymbol{\gamma}$, and $\vec{\pi} = \vec{p} - e\vec{A}$. The second term $-i e^{T} \partial_t e^{-T}$ introduces two fundamental challenges: it disrupts the block-diagonal structure and, as shown in~\cite{Goldman1977}, can lead to gauge-dependent energy expectation values. Conventional FW treatments avoid these issues by assuming static fields ($\partial_t \vec{A} = 0$), but this restriction limits their applicability to time-dependent scenarios.

In contrast, our approach achieves three key advantages while maintaining gauge consistency. First, we derive the relativistic Hamiltonian directly from Dirac's equation without iterative transformations, preserving all $\partial_t\vec{A}$ terms through the gauge-invariant identity $\vec{\nabla A}_0 = - (\vec{E} + \partial_t \vec{A})$. Second, we naturally obtain terms proportional to $(\vec{\sigma}\cdot\vec{\pi})^4$ and nested commutators that would require multiple FW transformations in conventional approaches, as explicitly evaluated in Appendix~C. Third, our formulation maintains consistency for both time-dependent fields and non-uniform magnetic fields, avoiding the gauge ambiguities inherent in FW-based methods while capturing all $\mathcal{O}(v^2/c^2)$ relativistic corrections.

The ability of our formulation to treat both uniform and non-uniform electromagnetic fields, as well as time-dependent configurations, demonstrates its wide applicability and distinguishes it from traditional FW-based methods.

The resulting Hamiltonian allows for treatment of both uniform and non-uniform magnetic fields, which are relevant in Penning traps, storage rings, and astrophysical environments. For example, the muon storage ring experiment~\cite{Farley1979} relies on a radial magnetic field gradient for vertical focusing, which introduces corrections to the anomalous precession frequency $\omega_s$. Similarly, although Penning trap experiments such as~\cite{Sturm2014,Heisse2023} are performed in highly uniform magnetic fields, there exists a tiny residual magnetic field inhomogeneity that leads to systematic errors. Our Hamiltonian, through terms like $\nabla_B^2$ and $\nabla_B \cdot \vec{p}$, offers a unique advantage in modeling these effects from the first principles, potentially reducing such uncertainties in high-precision $g$-factor measurements.

We apply our Hamiltonian to hydrogen-like atoms and compute the relativistic correction to the Land\'e $g$-factor ($g_J$) for arbitrary quantum numbers $n$, $j$, and $m_j$. Correction of order $\alpha^2$ introduces relativistic $m_j^2$-dependent terms, absent in conventional treatments, and matches the known results of Breit~\cite{Breit1928} in appropriate limits. Notably, this $m_j^2$-dependent relativistic correction is experimentally testable: as discussed in Section~\ref{subsec:penning}, Penning trap spectroscopy enables direct measurement of $g$-factor from the difference between magnetic sublevels (e.g.,  $m_j=3/2$ and  $m_j=1/2$), providing a sensitive test of our predictions.

Beyond our Hamiltonian framework, recent developments in the vector model of spinning particles have provided new insights into the interaction of spin with external fields, including gravitational and electromagnetic interactions~\cite{Deriglazov2017}. These models clarify the relationship between relativistic quantum mechanics and the Dirac equation, and highlight the role of spin-induced non-commutativity in the algebra of position variables~\cite{Deriglazov2016}. Furthermore, studies revisiting the anomalous magnetic moment of the electron using extended Dirac Hamiltonians show that radiative corrections affect both the spin and orbital $g$-factors~\cite{Awobode2002,Awobode2023}.

In this paper, we compute the first-order relativistic correction to the $g_J$-factor using our Hamiltonian (Equations~\ref{H00}--\ref{Hsl}) in hydrogen-like atoms. The $g_J$-factor includes both spin and orbital contributions, denoted by $g_s$ and $g_l$, respectively. Our relativistic correction at order $\alpha^2$ has the same negative sign as that found in~\cite{Sommerfield1957,Petermann1957}. Importantly, the expressions derived in equations ~\ref{g+} and \ref{g-} are valid for arbitrary $n$, $j$, and $m_j$, and reduce to the Breit results~\cite{Breit1928} when expanded to $\mathcal{O}(\alpha^2)$.

Corrections to the $g$-factor fall into two categories: quantum field theory (QFT) and relativistic effects. QFT corrections, such as the Schwinger term, stem from virtual particle loops and are intrinsic to the lepton. Relativistic corrections arise from solutions to the Dirac equation in external fields and depend on motion ($p^2$) and the structure of the field ($\nabla_B$). Although QFT effects dominate high-precision $g$-2 measurements since the relativistic effects cancel out, relativistic contributions become significant in high-$Z$ bound systems, relativistic scattering, and strong magnetic fields.

Equations~\ref{H00}--\ref{Hsl} decompose the total Hamiltonian into kinetic, orbital, spin, and spin-orbit terms. In hydrogen-like atoms, we relate the relativistic corrections in Eqs.\ref{Hs} and \ref{Hl} to the expectation value of the kinetic energy, and use them to compute $g_J$. In the absence of a magnetic field, the relativistic energy shift depends on both the principal quantum number $n$ and the total angular momentum quantum number $j$, but in the Zeeman effect, it also depends on $j$ and $m_j^2$, as shown in Eqs.~\ref{g+} and~\ref{g-}. Notably, the ground state exhibits the largest shift of order $\alpha^2/3$.

The structure of the paper is as follows: Section~2 introduces the Pauli equation and derives $g_s = 2$ from the Dirac equation. Section~3 derives the full relativistic Hamiltonian and highlights key operator expansions. Section~4 applies the Hamiltonian to compute the relativistic correction to $g_J$ in hydrogen-like atoms, recovering known results in specific limits. Section~5 discusses broader implications, including Penning traps, muonic atoms, and relativistic scattering. Conclusions follow in Section~6. Appendix A collects fundamental definitions and mathematical identities. Appendices B and C present detailed derivations of the $(\vec{\sigma} \cdot \vec{\pi})^4$ term and the double commutator structure, respectively.

\section{Pauli Equation for the Spin-Magnetic Interaction}

The standard Schrödinger equation does not account for the intrinsic spin of the electron or its interaction with a magnetic field. To incorporate these effects, the equation can be extended using the Pauli equation, which operates on a two-component spinor $\chi$. In the absence of external fields or interactions, the free-particle equation is given by:
\begin{equation}\label{2.1}
\frac{p^2}{2m}\chi = E\chi
\end{equation}
Here, $p^2 = \vec{p} \cdot \vec{p}$, and the kinetic energy operator is implicitly multiplied by the identity matrix $2 \times 2$.  Applying the Pauli identity (see Appendix~\ref{app:definitions}), we may rewrite equation (\ref{2.1}) as:
\begin{equation}\label{2.3}
\frac{(\vec{\sigma} \cdot \vec{p})(\vec{\sigma} \cdot \vec{p})}{2m}\chi = E\chi
\end{equation}
The matrices $\sigma_i$ are the standard Pauli matrices (see Appendix~\ref{app:definitions} for explicit forms). In the presence of an electromagnetic field, the linear momentum $\vec{p}$ is replaced by the canonical momentum $\vec{\pi}$ (defined in Appendix A). Substituting this into (\ref{2.3}), we obtain:
\begin{equation}\label{2.51}
\frac{(\vec{\sigma} \cdot \vec{\pi})(\vec{\sigma} \cdot \vec{\pi})}{2m}\chi = E\chi
\end{equation}
Because $\vec{\pi}$ includes both the momentum operator and the position-dependent vector potential $\vec{A}(\vec{r})$, and because these do not commute, we must apply Pauli's identity again carefully. Doing so yields the following:
\begin{equation}\label{2.6}
(\vec{\sigma} \cdot \vec{\pi})^2 = \pi^2 - \frac{e\hbar}{c} \vec{\sigma} \cdot \vec{B}
\end{equation}
where $\vec{p} = -i\hbar \vec{\nabla}$ and $\vec{B} = \vec{\nabla} \times \vec{A}$ is the magnetic field. Throughout this paper, we adopt the following notational convention: we use $\vec{p}$ when the derivative operator acts on the wavefunction, and $\vec{\nabla}$ when it acts on external fields or potentials (i.e., anything other than the wavefunction). Substituting equation (\ref{2.6}) into (\ref{2.51}), we arrive at the Pauli equation in the presence of a magnetic field:
\begin{equation}\label{2.8}
\left( \frac{\pi^2}{2m} - \frac{e\hbar}{2m} \vec{\sigma} \cdot \vec{B} \right)\chi = E\chi
\end{equation}

The first term governs the kinetic energy and motion of the electron in the electromagnetic field. The second term represents the interaction of the spin magnetic moment with the magnetic field. The interaction energy term can be isolated as:
\begin{equation}\label{2.9}
U_{\text{Pauli}} = -\frac{e\hbar}{2m} \vec{\sigma} \cdot \vec{B} = -\vec{\mu}_s \cdot \vec{B}
\end{equation}
where $\vec{\mu}_s$ is the spin magnetic moment, defined in equation (\ref{spin-magnetic}), with $g_s = 2$ at tree level, as predicted by Dirac theory.

\section{Relativistic Correction to the Dirac-Pauli Equation}

In this section, we derive the Pauli equation from the Dirac equation in the presence of an electromagnetic field, retaining relativistic corrections of order \( v^2/c^2 \) . We focus on extracting physically meaningful terms in the resulting Hamiltonian, particularly those associated with corrections to the magnetic moment.

The interaction between a charged lepton and an external electromagnetic field is governed by the Dirac equation with a covariant derivative:
\begin{equation}\label{3.1}
(i\gamma^\mu \hbar D_\mu(x) - mc)\Psi(x) = 0,
\end{equation}
where the covariant derivative $D_\mu(x)$ is defined in Appendix A. This derivative introduces the four-vector potential $A_\mu(x)$ to ensure local $U(1)$ gauge invariance of the Dirac Lagrangian. The invariant mass $m$ is defined in Appendix A. The gamma matrices $\gamma^\mu$ in the chiral (Weyl) representation (see Appendix A for explicit forms) are used throughout this work, and the Dirac spinor is split into two components:
\begin{equation}\label{3.5}
\Psi(x) = \begin{pmatrix} \psi_L(x) \\ \psi_R(x) \end{pmatrix}.
\end{equation}

We assume $A_\mu(x)$ is slowly varying with time, and that the dominant time dependence of $\Psi(x)$ is captured by the ansatz:
\begin{equation}\label{3.6}
\Psi(x) = \begin{pmatrix} u_L \\ u_R \end{pmatrix} e^{-iEt/\hbar},
\end{equation}
we can decompose the Dirac equation into two coupled equations:
\begin{align}
\vec{\sigma} \cdot \vec{\pi} \, u_R &= \left( \frac{E - eA_0}{c} - mc \right) u_L,\label{3.7}
 \\
\vec{\sigma} \cdot \vec{\pi} \, u_L &= \left( \frac{E - eA_0}{c} + mc \right) u_R.\label{3.8}
\end{align}

Eliminating \( u_R \) from equation (\ref{3.8}) using (\ref{3.7}), we obtain the following:
\begin{equation}\label{3.9}
(\vec{\sigma} \cdot \vec{\pi}) \frac{c^2}{E - eA_0 + mc^2} (\vec{\sigma} \cdot \vec{\pi}) u_L = (E - eA_0 - mc^2) u_L.
\end{equation}
where we have used \( p_0 = E/c \). The energy can be expanded above \( m c^2 \) as:
\begin{equation}\label{3.10}
  E = m c^2 + E_R. 
\end{equation}
Equation \ref{3.9} then becomes in terms of \( E_R \):

\begin{equation}\label{3.11}
 (\vec{\sigma} \cdot \vec{\pi}) \frac{c^2}{E_R - A_0 + 2 m_0 c^2} (\vec{\sigma} \cdot \vec{\pi}) u_L = (E_R - e A_0) u_L. 
 \end{equation}
 
The non-relativistic limit is obtained when $E_R \approx 0$, under the assumption that $2m_0 c^2 \gg e A_0$. In this regime, the Dirac equation reduces to a Pauli-like form (see \ref{2.8}), where the zeroth-order spin $g$ factor takes the value $g_s = 2$ \cite{Kaku1993,Ryder1996}, in agreement with Dirac theory.

However, when these assumptions are relaxed and we consider energy expansions above the rest mass, the magnetic moment of the charged lepton deviates from the non-relativistic prediction. This deviation manifests itself as a correction to the factor $g$, reflecting relativistic effects not taken into account in the leading order.

Furthermore, suppression of the lower spinor component $u_R$ relative to the upper component $u_L$ can be estimated from Eq.~(3.8). It follows that $u_R$ is smaller than $u_L$ by a factor of order $\frac{v}{c}$, consistent with expansion in small velocities.
In this section, we will go beyond the zeroth order and include the first relativistic-order correction. Up to this order, \( u_L \) is not well normalized, and we encounter a probabilistic difficulty, as required by the normalization condition of Dirac spinors:
To ensure proper normalization, we define the corrected spinor \( U_L = \Omega u_L \), where
\begin{equation}\label{3.12}
\Omega \approx \left( 1 + \frac{1}{8m^2 c^2} (\vec{\sigma} \cdot \vec{\pi})(1 + y)^{-2} (\vec{\sigma} \cdot \vec{\pi}) \right),
\end{equation}
with \( y = \frac{E_R - eA_0}{2mc^2} \). By applying this normalization to equation (\ref{3.11}), and multiplying both sides from the left by \( \Omega^{-1} \), we obtain:
\begin{equation}\label{3.13}
H U_L = E_R U_L,
\end{equation}
where the Hamiltonian is:
\begin{align}\label{3.14}
H \approx \frac{(\vec{\sigma} \cdot \vec{\pi})^2}{2m} + eA_0 
&- \frac{(\vec{\sigma} \cdot \vec{\pi})(E_R - eA_0)(\vec{\sigma} \cdot \vec{\pi})}{4m^2 c^2}
- \frac{(\vec{\sigma} \cdot \vec{\pi})^4}{8m^3 c^2} + \frac{\{ (\vec{\sigma} \cdot \vec{\pi})^2, (E_R - eA_0) \}}{8m^2 c^2}.
\end{align}

To simplify further, we use the identity defined in equation (\ref{eq:anticommutation}) and obtain the final simplified Hamiltonian:
\begin{equation}\label{3.16}
H \approx \frac{(\vec{\sigma} \cdot \vec{\pi})^2}{2m} + eA_0 - \frac{(\vec{\sigma} \cdot \vec{\pi})^4}{8m^3 c^2} + \frac{[\vec{\sigma} \cdot \vec{\pi}, [\vec{\sigma} \cdot \vec{\pi}, (E_R - eA_0)]]}{8m^2 c^2}.
\end{equation}
The detailed calculations of the third and fourth terms in equation (\ref{3.16}) are presented in Appendix B and C, respectively. Substituting equations (\ref{B-3}), (\ref{B-6}), and (\ref{C-4}) into equation (\ref{3.16}), we obtain:
\begin{equation}\label{H}
    H = H_0 + H_l + H_s + H_{sl},
\end{equation}
where
\begin{align}
  H_0 &\approx  \frac{p^2}{2m} + eA_0 + \frac{e^2}{2mc^2} A^2 - \frac{p^4}{8m^3 c^2} - \frac{e\hbar^2}{8m^2 c^2} \vec{\nabla} \cdot \vec{E}, \label{H00} \\
  H_l &\approx - \frac{e}{mc} \left(1 + \frac{\hbar^2}{4m^2 c^2} \nabla_A^2 + \frac{i\hbar}{2m^2 c^2} (\vec{\nabla}_A \cdot \vec{p}) - \frac{p^2}{2m^2 c^2} \right)(\vec{A} \cdot \vec{p}), \label{Hl} \\
  H_s &\approx - \frac{e\hbar}{2mc} \left(1 - \frac{p^2}{2m^2 c^2} + \frac{\hbar^2}{4m^2 c^2} \nabla_B^2 + \frac{i\hbar}{2m^2 c^2} (\vec{\nabla}_B \cdot \vec{p}) \right)(\vec{\sigma} \cdot \vec{B}) \nonumber \\
  &\quad - \frac{e^2 \hbar}{4m^2 c^3} \vec{\sigma} \cdot (\vec{A} \times \vec{E}) - \frac{e^2 \hbar}{4m^2 c^3} \vec{\sigma} \cdot \left( \vec{A} \times \frac{\partial \vec{A}}{\partial t} \right), \label{Hs} \\
  H_{sl} &\approx - \frac{e\hbar}{4m^2 c^2} \vec{\sigma} \cdot (\vec{E} \times \vec{p}) - \frac{e\hbar}{4m^2 c^2} \vec{\sigma} \cdot \left( \frac{\partial \vec{A}}{\partial t} \times \vec{p} \right). \label{Hsl}
\end{align}

Based on the type of interaction, the Hamiltonian is decomposed into four components. The term $H_0$ includes kinetic and potential energies along with their leading relativistic corrections. $H_l$ and $H_s$ describe the interaction of the orbital and spin magnetic moments with external fields respectively, including their relativistic corrections. $H_{sl}$ represents the spin-orbit interaction.

Starting with equation (\ref{H00}), the first and second terms correspond to classical kinetic and potential energies, respectively. The third term is the diamagnetic term, which is typically negligible for hydrogen-like atoms. The fourth term is the first relativistic correction to kinetic energy,  which arises from the relativistic energy expansion given in Appendix A. The last term, Darwin's term, relates to charge density via $\vec{\nabla} \cdot \vec{E} = \rho/\epsilon_0$, and contributes only to $S$ states of hydrogen atoms. In equation \ref{Hl}, the term $\vec{A} \cdot \vec{p}$ represents magnetic interaction with orbital motion, and the bracketed terms include relativistic corrections.

For equation (\ref{Hs}), the leading term describes spin-magnetic field interaction. The subleading terms are relativistic corrections. The second line contains interactions between spin and effective magnetic fields generated by motion of charged lepton through $\vec{E}$ or time-varying $\vec{A}$.

Equation~\eqref{Hsl} describes the spin-orbit interaction. Notably, the first term reproduces both Larmor and Thomas precession contributions: the spin interacting with an effective magnetic field and the precession of the spin vector, respectively.

The generality of the Hamiltonian in equations \ref{H}--\ref{Hsl} makes it suitable for both scattering and bound-state problems, including hydrogen-like atoms in external magnetic fields. However, the general Hamiltonian is valid under certain conditions: (1) the magnetic field $B$ must be less than $10^9$~T to avoid pair production; (2) the velocity $v$ must satisfy $v/c < 0.3$ to ensure that higher-order corrections beyond $\mathcal{O}(v^2/c^2)$ remain negligible. For bound systems where $v/c \approx Z\alpha$, this restricts the applicability to systems with $Z \lesssim 30$. Under these constraints, the truncation scheme used remains consistent and captures all dominant corrections.

The Hamiltonian in equations.~(\ref{H})--(\ref{Hsl}) is derived for charged leptons via minimal coupling, as shown in equation~(\ref{eq:covarient}). For neutral leptons (e.g., neutrinos) with \( e = 0 \), all electromagnetic interaction terms vanish, including \( H_l \), \( H_{sl} \), and the charge-dependent components of \( H_0 \) and \( H_s \). However, if a neutral lepton possesses a nonzero anomalous magnetic moment \( \kappa \), an additional Pauli term, \( \frac{\kappa}{4} \sigma^{\mu\nu} F_{\mu\nu} \Psi \), must be added to the right-hand side of the Dirac equation~(\ref{3.1}). Here, \( F_{\mu\nu} = \partial_\mu A_\nu - \partial_\nu A_\mu \) is the electromagnetic field strength tensor, and \( \sigma^{\mu\nu} = \frac{i}{2} [\gamma^\mu, \gamma^\nu] \) is the antisymmetric combination of gamma matrices. This leads to the following non-relativistic Hamiltonian:
\[
H = \frac{p^2}{2m} - \frac{p^4}{8m^3c^2} - \frac{\kappa \hbar}{2mc} \vec{\sigma} \cdot \vec{B} + \frac{\kappa \hbar}{4m^2c^2} \vec{\sigma} \cdot (\vec{E} \times \vec{p}),
\]
which describes spin–field coupling and its leading relativistic correction. Neutral leptons exhibit no orbital interaction with the magnetic field due to the absence of charge. In the Standard Model, \( \kappa = 0 \); even in extensions where \( \kappa \neq 0 \), its value is extremely small, and the relativistic correction further suppresses its already negligible contribution. Additional terms can arise in the presence of an anomalous magnetic moment, such as a coupling of the electric field to a shifted momentum operator \( \vec{p}\,' = \vec{p} + \frac{\kappa}{2} \vec{B} \). This yields a scalar interaction term \( \vec{p}\,' \cdot \vec{E} \), which arises from the anomalous magnetic moment and represents an additional spin-dependent interaction with the external electric field. In our formalism, further gradient terms involving \( \vec{B} \), such as \( \nabla^2 \vec{B} \), arise naturally through nested commutators in the Hamiltonian expansion. These structures are present in the charged case due to minimal coupling and may also appear in the neutral case through the Pauli term. All such terms are consistently suppressed by \( \kappa \) and \( v^2/c^2 \), preserving the order of approximation.

More generally, our formalism applies to any fermion interacting with electromagnetic fields within the framework of QED. For quarks, the same Hamiltonian structure applies, with the electric charge \( e \) replaced by the appropriate fractional charge \( Q_q e \), where \( Q_q = +\frac{2}{3} \) for up-type quarks and \( Q_q = -\frac{1}{3} \) for down-type quarks. This substitution preserves the general form of the Hamiltonian. Interactions beyond QED, such as those involving non-Abelian gauge fields in QCD, are beyond the scope of this work and are therefore not included in our analysis.

\section{Hydrogen-\hspace{0pt}like Atoms and Relativistic Corrections to the Magnetic Moment}
In this section, we apply the general Hamiltonian derived in equations (\ref{H})--(\ref{Hsl}) to the hydrogen-like atom, focusing on the first-order relativistic correction to the electron's magnetic moment arising from both its spin and orbital angular momenta. Our results match those obtained in \cite{Breit1928} up to order $\alpha^2$, confirming the validity of our approach for both ground and excited states.

Hydrogen-like atoms serve as a canonical example of systems with central spherical potentials, allowing transparent interpretation of each Hamiltonian term. In the presence of an external static magnetic field $\vec{B}$ aligned along the $z$-axis, we adopt the symmetric gauge for the vector potential:
\begin{equation}\label{4.1}
\vec{A} = \frac{1}{2} (\vec{B} \times \vec{r}).
\end{equation}

Under this gauge, and noting that certain operator terms vanish when acting on $\vec{A}$, the orbital interaction term and its relativistic correction simplify to:
\begin{equation}\label{4.2}
H_L \approx -\frac{e}{2mc} \left(1 - \frac{K}{mc^2}\right) \vec{B} \cdot \vec{L},
\end{equation}
where $K = \frac{p^2}{2m}$ is the kinetic energy operator and $\vec{L} = \vec{r} \times \vec{p}$ is the orbital angular momentum. The magnetic moment of the electron associated with its orbital motion is:
\begin{equation}\label{4.3}
\vec{\mu}_L = \frac{e\vec{L}}{2mc}.
\end{equation}

The spin-related Hamiltonian component from equation (3.22), assuming static and homogeneous $\vec{A}$ and $\vec{B}$ fields, becomes:
\begin{equation}\label{4.4}
H_S \approx -\frac{e\hbar}{2mc} \left[ (\vec{\sigma} \cdot \vec{B}) - \frac{p^2}{2m^2c^2} (\vec{\sigma} \cdot \vec{B}) + \frac{e}{2mc^2} \vec{\sigma} \cdot (\vec{A} \times \vec{E}) \right].
\end{equation}

Using the nuclear electric field for a hydrogen-like atom with atomic number $Z$:
\begin{equation}\label{4.5}
\vec{E} = -\frac{Ze}{4\pi \epsilon_0 r^2} \hat{r},
\end{equation}
we rewrite the third term of equation (\ref{4.4}) as:
\begin{equation}
e \vec{\sigma} \cdot (\vec{A} \times \vec{E}) = \frac{e}{2} (\vec{B} \times \vec{r}) \times \vec{E} = -U \cos \theta \, \vec{\sigma} \cdot \hat{r},
\end{equation}
where $U = -\frac{Ze^2}{4\pi \epsilon_0 r}$ and $\theta$ is the angle between $\vec{B}$ and $\vec{r}$.

Combining the orbital and spin parts of the Hamiltonian, we express:
\begin{equation}
H_{\sigma L} \approx H^0_{\sigma L} + H^r_{\sigma L},
\end{equation}
where the leading (non-relativistic) and relativistic Hamiltonians are given respectively by :
\begin{align}
H^0_{\sigma L} &\approx -\frac{e}{2mc} \vec{B} \cdot (2\vec{S} + \vec{L}), \\
H^r_{\sigma L} &\approx -\frac{e}{2m^2c^3} \left( \vec{F} \cdot \vec{S} - K \, \vec{B} \cdot \vec{L} \right),\label{Hr}
\end{align}
with the field-dependent vector:
\begin{equation}
\vec{F} = -2K\vec{B} - \frac{U}{2} \vec{B} + \frac{U \cos \theta}{2} B \hat{r}.
\end{equation}

The interaction between spin and orbital moments with the magnetic field is conventionally expressed as:
\begin{equation}
g_J \vec{J} \cdot \vec{B} = (g_s \vec{S} + g_l \vec{L}) \cdot \vec{B},
\end{equation}
where $g_s$ and $g_l$ are the spin and orbital $g$-factors, respectively. The expectation value of $H^0$ corresponds to the non-relativistic interaction energy between the magnetic moment and the external magnetic field. Let us consider the unperturbed eigenstates of a hydrogen-like atom in the absence of an external magnetic field, denoted by $|n, j, m_j\rangle$, where $n$ is the principal quantum number, $j$ is the total angular momentum quantum number, and $m_j$ is its projection along the field direction. Using the Wigner-Eckart theorem and projection operators for an arbitrary vector operator $\vec{A}$, we compute expectation values in the $|n, j, m_j\rangle$ basis:
\begin{equation}\label{4.12}
\langle j', m_j' | \vec{A} | j, m_j \rangle = \frac{\langle j, m_j | \vec{A} \cdot \vec{J} | j, m_j \rangle}{j(j + 1) \hbar^2} \langle j', m_j' | \vec{J} | j, m_j \rangle.
\end{equation}

This leads to the well-known expression for the Landé $g$-factor:
\begin{equation}\label{4.13}
g^0_J = 1 + \zeta(j, l, s),
\end{equation}
with
\begin{equation}\label{4.14}
\zeta(j, l, s) = \frac{j(j + 1) - l(l + 1) + s(s + 1)}{2j(j + 1)}.
\end{equation}
We substituted the leading-order values $g_s = 2$ and $g_l = 1$, and used the following angular momentum identities:
\begin{align}\label{SdotJ}
\vec{S} \cdot \vec{J} &= \frac{1}{2}(J^2 + S^2 - L^2), \nonumber\\
\vec{L} \cdot \vec{J} &= \frac{1}{2}(J^2 - S^2 + L^2).
\end{align}
Here, $\vec{J} = \vec{L} + \vec{S}$ is the total angular momentum operator. The eigenvalues of the squared angular momentum operators $Q^2$, where $Q = J, L, S$, satisfy $Q^2\vert j,l,s \rangle=q(q+1)\hbar^2 \vert j,l,s \rangle$. 
In order to compute the first-order relativistic correction to $\langle H^r_{\sigma l} \rangle$, we evaluate the expectation value $\langle \vec{F} \cdot \vec{S} \rangle$ from equation~\ref{Hr} using the projection theorem stated in equation~\ref{4.12}, as follows:
\begin{align}\label{projectiontheorem11}
\langle j',m_j'|F_i | j'',m_{j''}\rangle\langle j'',m_{j''}| S_i|j,m_j \rangle
= \langle j',m_{j'}|\vec{F} \cdot \vec{J}|j,m_j \rangle\, 
\frac{\langle j,m_j|\vec{S} \cdot \vec{J}|j,m_j \rangle}{j(j+1)\hbar^2}.
\end{align}

Here, we have inserted the completeness relation 
$\sum_{j'',m_{j''}} | j'',m_{j''} \rangle \langle j'',m_{j''} | = 1$. 
If we take $\langle j', m_{j'}| = \langle j, m_j|$, only the $z$-component of $\vec{F} \cdot \vec{J}$ contributes, leading to:
\[
\langle j,m_j | \vec{F} \cdot \vec{J} | j,m_j \rangle = m_j \langle j,m_j | F_z | j,m_j \rangle.
\]
Thus, we obtain
\begin{equation}
\langle j, m_j | F_z | j, m_j \rangle = -\langle K \rangle B \left( 1 + \langle \cos^2 \theta \rangle \right),
\end{equation}
where we used $\langle U \cos^2 \theta \rangle = \langle U \rangle \langle \cos^2 \theta \rangle$, justified by the spherical symmetry of the potential, and the virial theorem: $\langle U \rangle = -2 \langle K \rangle$.

The expectation values $\langle K \rangle$ and $\langle U \rangle$ can be determined by either solving the Dirac equation directly or referring to the relativistic energy spectrum of the hydrogen atom, given by \cite{Bjorken1964}:
\begin{align}\label{EnoB}
E_n &= mc^2 \left[1 + \left(\frac{Z\alpha}{n - \left( j + \frac{1}{2} \right) + \sqrt{\left( j + \frac{1}{2} \right)^2 - Z^2 \alpha^2}} \right)^2 \right]^{-\frac{1}{2}} \nonumber \\
&\approx mc^2 \left( 1 - \frac{(Z\alpha)^2}{2n^2} - \frac{(Z\alpha)^4}{2n^3} \left( \frac{1}{j+1/2} - \frac{3}{4n} \right) + \cdots \right).
\end{align}

Here, $j = l \pm \tfrac{1}{2}$. The $\mathcal{O}(\alpha^2)$ term represents the combined contributions of $\langle U \rangle + \langle K \rangle$, while the $\mathcal{O}(\alpha^4)$ term arises from the Darwin term, spin-orbit interaction, and higher-order kinetic energy correction. Since we retain only $\mathcal{O}(\alpha^2)$ corrections, we approximate:
\[
E_n \approx mc^2 + \langle U \rangle + \langle K \rangle \approx mc^2 - \frac{1}{2} \frac{(Z\alpha)^2}{n^2} mc^2,
\]
and therefore,
\begin{equation}
\langle K \rangle = \frac{(Z\alpha)^2}{2n^2} mc^2.
\end{equation}

Next, we compute the angular average $\langle \cos^2 \theta \rangle$ using:
\begin{equation}\label{theta1}
\langle \cos^2 \theta \rangle = \int y_{lj}^{\pm, m_j}(r, \theta, \phi) \cos^2 \theta \, y_{lj}^{\pm, m_j}(r, \theta, \phi) \sin \theta \, d\theta \, d\phi,
\end{equation}
where the spin-angular wavefunctions for $j_{\pm} = l \pm \tfrac{1}{2}$ are given by \cite{Bjorken1964}:
\begin{align}\label{theta}
y_{lj_{\pm}}^{\pm, m_{j_{\pm}}}(r, \theta, \phi) &= \pm \sqrt{ \frac{l \pm m_{j_{\pm}} + 1/2}{2l + 1} } Y_l^{m_{j_{\pm}} - 1/2} \begin{pmatrix} 1 \\ 0 \end{pmatrix} + \sqrt{ \frac{l \mp m_{j_{\pm}} + 1/2}{2l + 1} } Y_l^{m_{j_{\pm}} + 1/2} \begin{pmatrix} 0 \\ 1 \end{pmatrix}.
\end{align}

Evaluating this leads to the identity:
\begin{equation}
\langle \cos^2 \theta \rangle = \frac{j_{\pm}^2 + j_{\pm} - m_{j_{\pm}}^2}{2j_{\pm}(j_{\pm}+ 1)}.
\end{equation}

To evaluate the relativistic correction to the Landé $g$-factor, we calculate the expectation value of the interaction Hamiltonian,
\begin{equation}\label{Hzeman}
H_{\sigma l} = -\frac{e\hbar}{2mc} g_J m_j B,
\end{equation}
by matching it to the operator expectation values derived from equation (\ref{Hr}). This leads to the following general expressions for the relativistic-corrected Landé $g$-factor for arbitrary $j_{\pm} = l \pm \frac{1}{2}$ and $m_{j_{\pm}}$:

For $j_+ = l + \frac{1}{2}$:
\begin{equation}\label{g+}
g_{J+} = g^0_{J+} \left( 1 - \frac{Z^2 \alpha^2}{2n^2} \cdot \frac{j_+(4j_+ + 1)(j_+ + 1) - m_{j_+}^2}{2j_+(j_+ + 1)(2j_+ + 1)} \right),
\end{equation}

For $j_- = l - \frac{1}{2}$:
\begin{equation}\label{g-}
g_{J-} = g^0_{J-} \left( 1 - \frac{Z^2 \alpha^2}{2n^2} \cdot \frac{j_-(4j_- + 3)(j_- + 1) + m_{j_-}^2}{2j_-(j_- + 1)(2j_- + 1)} \right),
\end{equation}

The leading-order (non-relativistic) Landé $g$-factors used in the above equations are:
\begin{align}
g^0_{J+} = \frac{2j_+ + 1}{2j_+},~~~~~~~~
g^0_{J-} = \frac{2j_- + 1}{2(j_- + 1)}.
\end{align}
The energy levels of a hydrogen-like atom in the presence of an external magnetic field are split according to the allowed values of $m_j$, ranging from $-j$ to $+j$, in accordance with the Zeeman effect. However, these levels experience additional shifts due to relativistic corrections of order $\alpha^2$ to the Landé $g$-factor, as shown in equations.~\eqref{g+} and \eqref{g-}.

The magnitude of the relativistic energy shift depends on the value of $m_j$. For states with $j_+ = l + \tfrac{1}{2}$, the energy shift is minimal when $m_{j_+} = \pm j_+$ and maximal when $m_{j_+} = \pm \tfrac{1}{2}$. Conversely, for $j_{-} = l - \tfrac{1}{2}$, the largest shift occurs at $m_{j_-} = \pm j_-$ and the smallest at $m_{j_-} = \pm \tfrac{1}{2}$.

This variation arises because the relativistic correction terms are proportional to $m_{j_{\pm}}^2$, and the structure of the expressions of the $g$ factor for $j_{\pm} = l \pm \tfrac{1}{2}$ reverses the dependence of the correction magnitude on $m_j$. As a result, the relativistic correction modifies the Zeeman splitting asymmetrically, depending on the total angular-momentum coupling scheme.

\begin{figure}[ht]
\centering
\includegraphics[width=0.75\textwidth]{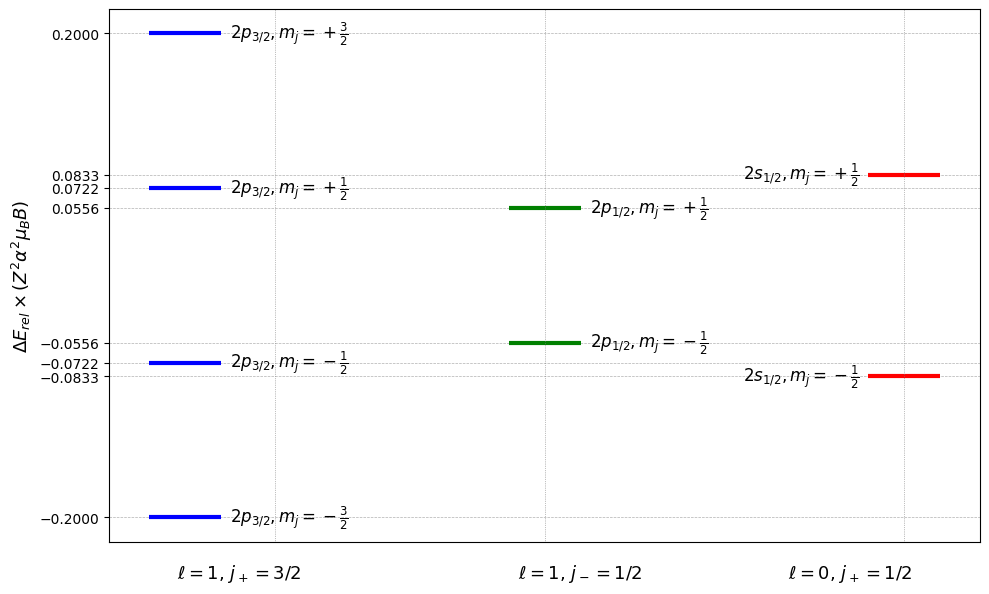}
\caption{First-order relativistic correction $\Delta E_{\text{rel}}$ to the Zeeman energy level for $n = 2$ hydrogen-like states, plotted as a function of $m_j$ for each fine-structure multiplet: $2p_{3/2}$ (blue), $2p_{1/2}$ (green), and $2s_{1/2}$ (red). The energy scale is normalized by $Z^2\alpha^2 \mu_B B$, where $\mu_B = \frac{e\hbar}{2mc} \approx 5.788 \times 10^{-5}$ eV/T is the Bohr magneton. The energy shift is proportional to $\Delta E_{\text{rel}} Z^2 \mu_B B$, showing how relativistic effects scale with atomic number $Z^2$. The results are based on equations ~\eqref{Hzeman}, \eqref{g+}, and \eqref{g-}. Spectroscopic notation $nl_j$ is used to denote the principal quantum number $n$, orbital angular momentum $l$ ($s=0$, $p=1$), and total angular momentum $j_{\pm} = l \pm 1/2$}
\label{fig:relshift}
\end{figure}

In order to obtain a rough estimate of the relativistic energy shift, we use the generalized expressions given in equations~\eqref{g+} and \eqref{g-}. Figure~\ref{fig:relshift} illustrates the first-order relativistic corrections to the Zeeman energy levels for the states $n = 2$, covering all allowed values of $j$, $l$, and $m_j$, for an arbitrary value of the atomic number $Z$.  The shift scales with $Z^2$, which becomes particularly relevant for heavier hydrogen-like ions.  This visual representation highlights the non-uniform splitting pattern arising from relativistic effects, particularly their dependence on $j$ and $m_j$.

In the special case where $m_{j_{\pm}} = \pm j$, equations \eqref{g+} and \eqref{g-} simplify to:
\begin{align}
g_{J+} &= g^0_{J+} \left( 1 - \frac{Z^2 \alpha^2}{4n^2} \cdot \frac{2j_+ + 1}{j_+ + 1} \right),
\\
g_{J-} &= g^0_{J-} \left( 1 - \frac{Z^2 \alpha^2}{4n^2} \cdot \frac{2j_- + 3}{j_- + 1} \right).
\end{align}

The corresponding energy shifts due to these relativistic corrections are:
\begin{align}
\frac{E_+ - E_+^0}{E_+^0} &\approx - \frac{Z^2 \alpha^2}{(l_{\text{max}} + 1)(2l_{\text{max}} + 3)},\label{Ep} \\
\frac{E_- - E_-^0}{E_-^0} &\approx - \frac{Z^2 \alpha^2}{(l_{\text{max}} + 1)(2l_{\text{max}} + 1)}\label{Em}.
\end{align}

Here, $E_0^{\pm}$ denotes the non-relativistic energy level for total angular momentum $j_{\pm} = l_{\text{max}} \pm 1$. As shown in equation~\eqref{Ep} and \eqref{Em}, the magnitude of relativistic corrections decreases with increasing principal quantum number $n$. This trend reflects the physical fact that the electron's velocity decreases at higher energy levels, reducing the significance of relativistic effects.

Consequently, the relativistic energy shift—of order $Z^2 \alpha^2 / 3$—is largest in the ground state and decreases with increasing $n$. In particular, for the excited-state configuration $n = l_{\text{max}} + 1$ and the ground-state configuration $j = 1/2$, the expression given in equation~\eqref{g+} precisely matches the $\mathcal{O}(\alpha^2)$ expansion of the two results originally derived by Breit \cite{Breit1928}.

The magnetic moment of the electron in terms of the Landé $g$-factor is then given by:
\begin{equation}\label{mu}
\vec{\mu}_e = \frac{e g_J}{2mc} \vec{J}.
\end{equation}

This magnetic moment includes contributions from both the spin and orbital angular momenta. In particular, the spin contribution is present in all states, whereas the orbital contribution vanishes for states with $l = 0$.

\section{Quantum Field Theory and Relativistic Corrections}

The magnetic moment of the electron is influenced by two distinct types of corrections: quantum field theory (QFT) corrections and relativistic corrections. These corrections arise from different physical mechanisms and play complementary roles in determining the electron's effective $g$-factor.

QFT corrections, such as the Schwinger term $a_{\text{QFT}} = \frac{\alpha}{2\pi}$, stem from higher-order loop processes involving virtual particles, including photons and leptons. These corrections are intrinsic properties of the particle and are independent of external fields. The anomalous magnetic moment $a_{\ell} = \frac{g_\ell - 2}{2}$ can be written as a power series in $\alpha/\pi$:
\begin{equation}
a_{\text{QED}} = c_1 \left(\frac{\alpha}{\pi}\right) + c_2 \left(\frac{\alpha}{\pi}\right)^2 + c_3 \left(\frac{\alpha}{\pi}\right)^3 + \cdots,
\end{equation}
where, for a free electron, $c_1 = \frac{1}{2}$. Higher-order QFT corrections (e.g., \(\alpha^2, \alpha^3, \dots\)) are also intrinsic and are computed using perturbative quantum electrodynamics (QED) \cite{Aoyama2012}.

Relativistic corrections, by contrast, emerge from solving the Dirac equation in external electromagnetic fields. These corrections depend on the particle's motion (via $\vec{p}$) and the spatial structure of the field (e.g., gradients $\nabla \vec{B}$). For example, in hydrogen-like atoms, the relativistic correction to the $g$-factor is proportional to $\alpha^2 Z^2$, where $Z$ is the atomic number of the nucleus. This correction becomes significant for highly charged ions~\cite{Shabaev2015,Lindroth1995}. In scattering systems, relativistic corrections arise due to the interaction between the electron and external electromagnetic fields. For an external electric field, the correction is typically of order $\alpha^4$, while for an external magnetic field, the correction can be of order $\alpha^2$, since magnetic interactions couple more directly to the electron's spin and orbital motion at a lower order in $\alpha$.

The quantum field theory (QFT) and relativistic corrections to the $g$-factor of a lepton originate from distinct physical mechanisms, as schematically illustrated in Figure ~\ref{fig:qft_flowchart}. The left branch represents QFT loop corrections, while the right branch shows relativistic corrections derived from solving the Dirac equation in external fields. Although the origins of these two effects differ, they play complementary roles in determining the effective $g$-factor observed in precision experiments.

\begin{figure}[H]
\centering
\begin{tikzpicture}[
    node distance=1.5cm,
    box/.style={draw, rounded corners, align=center, minimum width=2.5cm},
    arrow/.style={-{Latex[length=3mm]}, thick}]
% QFT branch (intrinsic)
    \node[box, fill=blue!10] (qft) {QFT Contributions\\ (Field-independent)};
    \node[box, below=of qft, fill=blue!10] (schwinger) {$\mathcal{O}(\alpha)$: Schwinger term\\ $\alpha/2\pi$};
    \node[box, below=of schwinger, fill=blue!10] (vacpol) {Self energy\\ (Vertex correction )};
    
    % Relativistic branch (field-dependent)
    \node[box, right=of qft, xshift=3cm, fill=green!10] (rel) {Relativistic Corrections\\ (Field-dependent)};
    \node[box, below=of rel, fill=green!10] (p2) {$\mathcal{O}(\alpha^2)$\\ (First-order correction)};
    \node[box, below=of p2, fill=green!10] (p4) {Kinetic and field correction\\ ($p^2/2m^2c^2$)\\ $v^2/c^2 \sim \alpha^2$};
    \node[box, below=of p4, fill=red!10] (mj) {Relativistic $m_j^2$-dependent term\\(This work)};
    
    % Arrows
    \draw[arrow] (qft) -- (schwinger);
    \draw[arrow] (schwinger) -- (vacpol);
    \draw[arrow] (rel) -- (p2);
    \draw[arrow] (p2) -- (p4);
    \draw[arrow] (p4) -- (mj);
    
    % Highlight region
    \node[draw=red, dashed, very thick, inner sep=8pt, fit=(schwinger) (vacpol) (p4) (mj), 
         label={[red, align=center]above:{Dominant in strong fields\\ or high-$Z$ ions}}] {};

\end{tikzpicture}
\caption[QFT and relativistic contributions to the $g$-factor]{
Decomposition of $g$-factor corrections.  
\textbf{Left (blue):} Leading QFT term ($\alpha/2\pi$) from electron self-energy via virtual photon exchange.  
\textbf{Right (green):} Relativistic $\mathcal{O}(\alpha^2)$ contributions from expectation values of kinetic and electromagnetic field operators, yielding the $m_j^2$-dependent term derived in this work.   
\textbf{Red box:} Dominant corrections in high-$Z$ or strong-field regimes.}
\label{fig:qft_flowchart}
\end{figure}

In high-precision $g\!-\!2$ experiments, such as those measuring the muon or electron anomaly, relativistic contributions are carefully considered and systematically eliminated. The anomalous frequency $\omega_a$ is defined as the difference between the spin precession frequency $\omega_s$ and the cyclotron frequency $\omega_c$:
\begin{align}\label{frequency}
\omega_s &= \frac{eB}{2mc} \left[ 2 \left(1 - \frac{p^2}{2m^2 c^2} \right) + 2a_{\ell} \right], \nonumber \\
\omega_c &= \frac{eB}{mc} \left(1 - \frac{p^2}{2m^2 c^2} \right),
\end{align}
yielding:
\begin{equation}
a_{\ell} = \frac{mc}{eB} (\omega_s - \omega_c).
\end{equation}
Here, the relativistic corrections to the spin and orbital interactions, derived in equations ~\eqref{Hl} and \eqref{Hs}, contribute to the frequency expressions in equation~\eqref{frequency}, and must be subtracted to isolate the pure QFT contribution to $a_{\ell}$. This experimental setup isolates the QFT contribution by canceling relativistic effects. The total correction to the $g$-factor can be written as an effective combination of QFT and relativistic terms:
\begin{equation}
g_{\text{eff}} = 2 \left(1 + a_{\text{QFT}} + a_{\text{rel}} \right),
\end{equation}
where $a_{\text{rel}}$ is the relativistic correction given by
\begin{align}\label{a_rel}
 a_{\text{rel}} = \frac{1}{2} \left( g_{J_{\pm}} - g_{J_{\pm}}^0 \right) =
\begin{cases} 
-\dfrac{Z^2\alpha^2}{4n^2} \cdot \dfrac{j_+(4j_+ + 1)(j_+ + 1) - m_{j_+}^2}{4j_+^2(j_+ + 1)}, \\[10pt]
-\dfrac{Z^2\alpha^2}{4n^2} \cdot \dfrac{j_-(4j_- + 3)(j_- + 1) + m_{j_-}^2}{4j_-(j_- + 1)^2}.
\end{cases}
\end{align}

In hydrogen-like atoms, $a_{\text{rel}} \sim \alpha^2 Z^2$ depends on the atomic number $Z$, while in scattering systems it depends on the kinematic regime and external field strength. To incorporate both QFT and relativistic corrections into the $g$-factor of hydrogen-like atoms, we add the QED Schwinger term to the relativistic result:
\begin{equation}
g_{J} = g_{J_{\pm}} + 2a_{\text{QED}} \approx g_{J_{\pm}} + \frac{\alpha}{\pi},
\end{equation}
where $g_{J_{\pm}}$ are the relativistic Landé $g$-factors defined in Eqs.~\eqref{g+} and~\eqref{g-}.

This combination is especially important in highly charged ions, where relativistic corrections become comparable to QFT corrections. The condition for this balance is:
\begin{equation}
Z \sim \frac{1}{\sqrt{\alpha}} \approx 11.7,
\end{equation}
indicating that relativistic effects dominate for $Z \gtrsim 10$.

In summary, QFT corrections dominate in weak-field and low-$Z$ systems, while relativistic corrections become essential in strong-field or high-velocity regimes, including:
\begin{itemize}
  \item Highly charged ions, where strong nuclear Coulomb fields cause relativistic effects to compete with QFT contributions~\cite{Shabaev2002},
  \item Extreme magnetic field environments, such as neutron stars and magnetars, where relativistic effects can dominate and perturbative QED approaches may break down~\cite{Harding2006,Duncan1992},
  \item Scattering systems involving relativistic leptons interacting with external fields.
\end{itemize}
These corrections are complementary and must both be taken into account for a complete and accurate theoretical prediction of the electron’s magnetic moment in both bound and scattering systems.

\begin{figure}[H]
\centering
\includegraphics[width=0.75\textwidth]{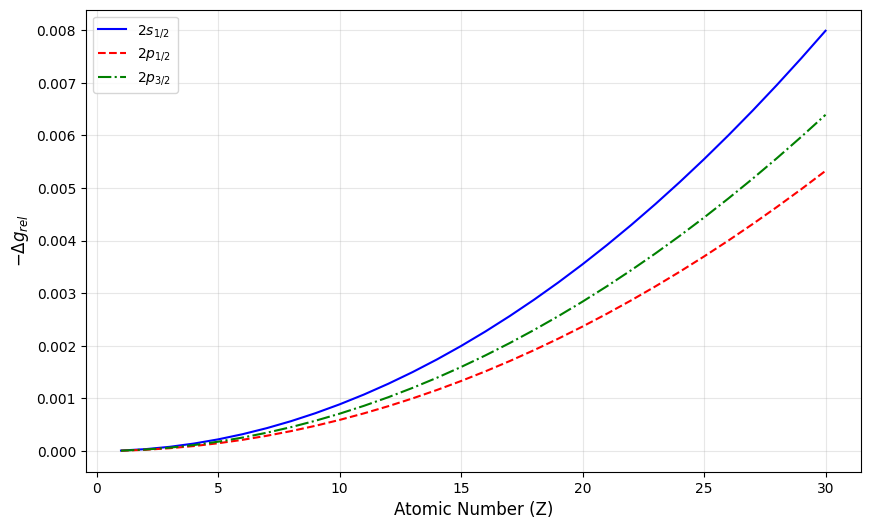}
\caption{
    Magnitude of the relativistic Land\'e $g$-factor shift ($-\Delta g_{\text{rel}}$) 
    for $2s_{1/2}$, $2p_{1/2}$ and $2p_{3/2}$ states in hydrogen-like atoms as a function of atomic number $Z$. The shift $\Delta g_{\text{rel}} = g_J - g_J^0$ is negative (reduction in $g$-factor), so $-\Delta g_{\text{rel}}$ is positive. 
    The $2s_{1/2}$ state have the largest correction, followed by $2p_{3/2}$ and then $2p_{1/2}$.  
    At $Z=20$, magnitudes align with Table~\ref{tab:gfactors} ($-\Delta g_{\text{rel}} \approx 0.0024$ and $0.0028$).}
\label{fig:figure3}
\end{figure}

To quantify the impact of relativistic corrections in leptons and atomic states, we present two key analyses. Table~\ref{tab:lepton_corrections} summarizes the order-of-magnitude relativistic corrections \( a_{\text{rel}} \) for electrons, muons, and tau leptons in hydrogen-like systems.\footnote{Tauonic atoms are hypothetical and are not expected to be observed directly in nature due to the extremely short lifetime of the tau lepton (\( \sim 10^{-13} \,\text{s} \)), which decays before a bound state can form.} These values demonstrate how the corrections scale with both the lepton mass \( m_\ell \) and the nuclear charge \( Z \), highlighting the increasing relevance of relativistic effects for heavier leptons and higher atomic numbers. Complementing this, Table~\ref{tab:gfactors} compares non-relativistic and relativistic Land\'e \( g \)-factors for selected \( n = 2 \) states. This comparison, alongside Figure~\ref{fig:figure3}, illustrates the \( Z^2 \) dependence of the relativistic shift \( \Delta g_{\text{rel}} = g_J - g_J^0 \) and the associated \( m_j \)-splitting effects. 

Figure~\ref{fig:figure3} explicitly shows the magnitude of the relativistic shift $-\Delta g_{\text{rel}}$ (where $ \Delta g_{\text{rel}} < 0 $) as a function of the atomic number $ Z $ for the states $2s_{1/2}$, $ 2p_{1/2} $ and $ 2p_{3/2} $. The $2s_{1/2}$ state exhibits the largest relativistic shift due to its finite probability density ($|\psi(0)|^2 \neq 0$) which enhances the relativistic kinetic correction ($\langle (\vec{\sigma}\cdot\vec{\pi})^4 \rangle$) . For the $2p$ states, the $2p_{3/2}$ shift exceeds $2p_{1/2}$ because aligned spin and orbital angular momentum ($j=3/2$) enhance the relativistic correction, whereas anti-aligned ($j=1/2$) configurations reduce it. This $j$-dependence originates from the $m_j^2$ terms in equations (4.24)-(4.25), with all shifts scaling as $Z^2\alpha^2$ in the magnetic field.  At $ Z = 20 $, the shifts reach $ \sim 0.3\% $ (Table~\ref{tab:gfactors}). These corrections become significant ($ \backsim 0.5\% $) for $ Z \approx 30 $, which impacts precision spectroscopy in highly charged ions.

\begin{table}[h]
\centering
\caption{Relativistic corrections $a_{\text{rel}}$ to the lepton $g$-factor in hydrogen-like atoms. The dominant contribution comes from the ground state ($n=1$) and scales as $a_{\text{rel}} \sim Z^2\alpha^2 (m_\ell/m_e)$. For comparison, the QED correction is $a_{\text{QFT}} \approx \alpha/\pi \approx 0.0023$. The last column shows the ratio $a_{\text{rel}}/a_{\text{QFT}}$.}
\vspace{0.5em}
\label{tab:lepton_corrections}
\begin{tabular}{lccccc}
\hline
Lepton & Mass ($m_\ell$) & $a_{\text{rel}}$ ($Z=1$) & $a_{\text{rel}}$ ($Z=20$) & Dominant Term & $\frac{a_{\text{rel}}}{a_{\text{QFT}}}$ ($Z=20$) \\
\hline
$e^-$ & $m_e$ & $1.8 \times 10^{-5}$ & $7.2 \times 10^{-3}$ & $\frac{Z^2\alpha^2}{3}$ & $3.13$ \\
$\mu$ & $207 m_e$ & $3.7 \times 10^{-3}$ & $1.5$ & $\frac{Z^2\alpha^2}{3}\frac{m_\mu}{m_e}$ & $645$ \\
$\tau$ & $3477 m_e$ & $0.062$ & $24.8$ & $\frac{Z^2\alpha^2}{3}\frac{m_\tau}{m_e}$ & $10,668$ \\
\hline
\end{tabular}
\end{table}

\begin{table}[h]
\centering
\caption{Non-relativistic ($g_J^0$) and relativistic ($g_J$) Land\'e $g$-factors for $n=2$ states. The shift $\Delta g_{\text{rel}} = g_J - g_J^0$ scales as $Z^2$.}
\label{tab:gfactors}
\vspace{0.5em}
\begin{tabular}{lcccc}
\hline
State & $g_J^0$ (Non-rel.) & $g_J$ (Rel., $Z=20$) & $\Delta g_{\text{rel}}$ ($Z=20$) & $\Delta g_{\text{rel}}$ ($Z=1$) \\
\hline
$2s_{1/2}$ & 2.0000 & 1.99645 & $-0.00355$ & $-8.88\times10^{-6}$ \\
$2p_{1/2}$ & 0.6667 & 0.66430 & $-0.00237$ & $-5.92\times10^{-6}$ \\
$2p_{3/2}$ & 1.3333 & 1.33049 & $-0.00284$ & $-7.10\times10^{-6}$ \\
\hline
\end{tabular}
\end{table}

Together, these tables and figures provide a consistent numerical framework for evaluating relativistic contributions to magnetic moments. Table~\ref{tab:lepton_corrections} emphasizes the mass hierarchy $ \tau > \mu > e^- $ of relativistic effects, while Table~\ref{tab:gfactors} highlights state-dependent corrections growing quadratically with $ Z $. The trends align with the structure and splitting behavior in Figure~\ref{fig:relshift}.

\subsection{Detailed Comparisons with Existing Literature}
\label{subsec:comparisons}

Our relativistic correction agrees with and extends Breit's formalism~\cite{Breit1928}. For states with $n = l_{\text{max}} + 1$ and $m_j = \pm j$, our result reduces to the standard Breit expression at order $\mathcal{O}(\alpha^2)$, confirming consistency with earlier theory. More generally, our Hamiltonian introduces three extensions: (i) it incorporates explicit $m_j^2$ dependence for arbitrary quantum states, (ii) it includes spatially varying magnetic fields via gradient terms such as $\nabla \vec{B}$, and (iii) it retains full kinetic energy contributions through $p^2$ terms in the spin Hamiltonian $H_s$ (see Eq.~\ref{Hs}).

The bound-electron $g$-factor receives contributions from several physical effects, which can be expressed as:
\begin{equation}
g = 2 + \Delta g_{\text{rel}} + \Delta g_{\text{QED}} + g_N,
\end{equation}
where $\Delta g_{\text{rel}}$ denotes the first-order relativistic correction derived in this work (Eqs.~\ref{g+}--\ref{g-}), $\Delta g_{\text{QED}}$ contains quantum electrodynamics (QED) corrections and $g_N$ accounts for nuclear contributions\cite{Mohr1998}.
In bound-state QED, the $g$-factor correction takes the form:
\begin{equation}
\Delta g_{\text{QED}} = \underbrace{\frac{\alpha}{2\pi}}_{\text{Schwinger term}} + \underbrace{\frac{\alpha}{2\pi}\frac{(Z\alpha)^2}{6n^2}}_{\text{binding correction}} + \mathcal{O}(\alpha^2(Z\alpha)^2),
\end{equation}
where the first term represents the free-electron anomalous magnetic moment, and the second term accounts for binding effects in atomic systems. For high-$Z$ ions ($Z \gtrsim 10$), our relativistic correction scales as $\Delta g_{\text{rel}} \sim \alpha^2 Z^2$ and becomes comparable in magnitude to the QED binding term~\cite{Shabaev2015,Lindroth1995}.

Moreover, even the leading-order QED coefficient $c_1 = \frac{1}{2}$ is slightly modified in bound systems due to vacuum polarization and self-energy effects. These corrections scale as $Z \alpha^2$ and shift $c_1$ from its free-electron value. For example, $c_1 = 0.500349$ for $Z = 8$, as shown in~\cite{Pachucki2004, Pachucki2005}, based on higher-order QED calculations that include self-energy~\cite{Yerokhin2008} and vacuum polarization~\cite{Jentschura2009}.

Although our primary focus is on relativistic corrections, nuclear effects also contribute and are represented by $g_N$. These include:
\sloppy
\begin{itemize}
\item \textbf{Recoil corrections}, which account for the finite nuclear mass and scale as $\sim (Z\alpha)^2 \cdot (m_e/m_N)$~\cite{Malyshev2020};
\item \textbf{Finite nuclear size corrections}, arising from the non-pointlike charge distribution of the nucleus and scaling as $\sim (Z\alpha)^4 \cdot (R_N / \lambda_C)^2$, where $R_N$ is the rms nuclear charge radius and $\lambda_C = \hbar / mc$ is the electron's Compton wavelength~\cite{Glazov2002};
\item \textbf{Nuclear polarization}, with a contribution of order $\sim Z^2 \alpha^5$.
\end{itemize}
\fussy

For low- to moderate-$Z$ systems, these nuclear contributions remain small relative to $\Delta g_{\text{rel}}$ and $\Delta g_{\text{QED}}$, and their omission from our Hamiltonian-based framework is therefore justified within the targeted accuracy of this analysis.
\subsection{Other Possible Applications}
We have applied our generalized Hamiltonian to hydrogen-like atoms, its broader formulation enables application to other precision systems—particularly Penning trap experiments and muonic atoms.

The relativistic corrections derived in equations ~\eqref{g+} and \eqref{g-} include terms proportional to $m_j^2$, representing relativistic $m_j^2$-dependent shifts in the bound-electron $g$-factor. In Penning traps, the Hamiltonian provides a framework to test the predicted $m_j^2$-dependence of relativistic corrections through high-precision measurements of bound-electron $g$-factors. In muonic hydrogen atoms, the analysis parallels that of electronic hydrogen, but relativistic effects are enhanced by the muon's greater mass ($m_\mu \approx 207 m_e$), making such systems ideal for probing relativistic quantum electrodynamics in strong-field regimes.

\subsubsection{Applications to Penning Traps}
\label{subsec:penning}

The relativistic $m_j^2$-dependent shift can be directly tested in Penning trap experiments, which provide an ideal platform for isolating small corrections to the magnetic moment of charged particles.

A Penning trap confines a hydrogen-like ion using a strong, uniform magnetic field $\vec{B}$ and a static electric quadrupole potential. The ion undergoes circular motion in the transverse plane with a cyclotron frequency:
\[
\omega_c = \frac{q_{\text{ion}} B}{m_{\text{ion}}},
\]
where $q_{\text{ion}}$ and $m_{\text{ion}}$ are the ion's charge and mass, respectively. For a hydrogen-like ion, $q_{\text{ion}} = (Z - 1)e$. Simultaneously, the bound electron's spin precesses in the magnetic field at the Larmor frequency:
\[
\omega_L = \frac{g(m_j) e B}{2 m_e},
\]
where $m_e$ is the electron mass and $g(m_j)$ is the $g$-factor dependent on $m_j$. By measuring the ratio of these two frequencies, the effective bound-electron $g$-factor is obtained as:
\[
g(m_j) = 2 \left( \frac{\omega_L(m_j)}{\omega_c} \right) \left( \frac{m_e}{m_{\text{ion}}} \right) \left( \frac{q_{\text{ion}}}{e} \right).
\]

To test the predicted relativistic $m_j^2$- dependence, ions are prepared at different magnetic sublevels (e.g., $m_j = 3/2$ and $m_j = 1/2$). Measurement of the difference in the $g$ factors isolates the relativistic $m_j^2$ shift:
\[
\Delta g = g(m_j = 3/2) - g(m_j = 1/2),
\]
which cancels many nuclear structure and instrumental effects.

According to Eq.~\eqref{g+}, the relativistic correction for $j = 3/2$ states is of order $\Delta g \sim (Z^2 \alpha^2 / 90)$. For hydrogen-like carbon ($Z = 6$), this yields $\Delta g \approx 2.1 \times 10^{-5}$, corresponding to a Larmor frequency shift of approximately 600~Hz in a 1~T magnetic field. This level of precision is well within reach of modern Penning trap experiments, which can achieve relative uncertainties as small as $\delta g / g \sim 10^{-10}$~\cite{Diederich1999,Sturm2014,Heisse2023}. Thus, Penning trap spectroscopy offers a direct and highly sensitive test of the relativistic $m_j^2$-dependence in the bound-electron $g$-factor, as predicted by our relativistic Hamiltonian.

In addition, our generalized Hamiltonian $H_s$ (equation ~\eqref{Hs}) offers a key advantage in $g$-factor measurements within Penning traps by explicitly including magnetic field gradient terms such as $\nabla_B^2$ and $\nabla_B \cdot \vec{p}$. Real Penning traps often feature small field inhomogeneities—sometimes intentionally introduced for vertical focusing—which lead to systematic uncertainties. These were traditionally handled with phenomenological models, but our formalism allows such effects to be treated from first principles. By accounting for spatial variations of the magnetic field directly within the relativistic Hamiltonian, we improve the modeling accuracy for the electron’s magnetic interaction, potentially reducing one of the leading sources of uncertainty in precision $g$-factor experiments. While the present analysis focuses on hydrogen-like systems, the relativistic Hamiltonian derived here is applicable to many-electron ions when incorporated into many-body frameworks such as MCDHF or MBPT. For example, recent calculations on Na-like ions \cite{Rathi2022} confirm that relativistic corrections dominate in high-Z systems and are essential for interpreting high-precision Penning-trap measurements \cite{Sturm2014, Arapoglou2019, Sailer2022, Heisse2023}.

\subsubsection{Applications to Muonic Atoms}
\label{subsec:muonic}

The derived Hamiltonian (Equations \ref{H}--\ref{Hsl}) applies directly to muonic atoms, where relativistic effects are significantly amplified because the muon mass ($m_{\mu}\approx 207m_e$) reduces Bohr radii, increasing relativistic corrections by approximately $m_{\mu}/m_e\approx 207$ times. Although our current analysis focuses on these relativistic corrections to the magnetic moment and assumes a static proton, neglecting its recoil, a more comprehensive investigation of muonic hydrogen atoms presents crucial avenues for future work. In these exotic systems, the muon's significantly larger mass leads to a considerably smaller Bohr radius, causing it to orbit much closer to the proton. Consequently, effects such as proton recoil and its finite size, often negligible in electronic hydrogen, become highly pronounced and would necessitate careful consideration for a complete theoretical treatment. Furthermore, incorporating the spin-spin interaction between the muon and the proton, along with the proton's intrinsic magnetic moment, is essential for a full analysis of the system's energy levels and spectral properties, particularly given the insights from the proton radius puzzle \cite{Pohl2013, Gao2022}. These detailed extensions, which include proton dynamics and internal structure, represent important directions for building upon the relativistic Hamiltonian derived in this work.

\section{Conclusion}  

The interaction between the spin magnetic moment and the magnetic field is explicitly revealed by decomposing the four-component Dirac equation into the Pauli-Dirac equation. The non-relativistic limit of this decomposition yields the tree-level value $g_s = 2$, as predicted by Dirac theory. By including the first-order relativistic correction, a deviation from $g_s = 2$ becomes apparent. We derive a general Hamiltonian up to the first-order relativistic correction, which lays the groundwork for further investigations into various physical quantum systems, such as bound and scattering states. In this paper, we apply this general Hamiltonian to hydrogen-like atoms under the influence of a uniform magnetic field. The first-order relativistic correction to the Landé g-factor of the electron, related to its magnetic moment by equation \ref{mu}, is derived. A negative energy shift of order $\alpha^2$ arises as a consequence of the first relativistic correction. This energy shift depends on the quantum numbers $n$, $j$, and the square of $m_j$. On the other hand, the negative energy shift for hydrogen-like atoms in the absence of an external magnetic field depends on $n$ and $j$, as shown by equation \ref{EnoB}. By selecting specific orbits, such as $n = l_{\text{max}} + 1$, $m_j = \pm j$, and $j = l_{\text{max}} + 1/2$, the result in equation \ref{g+} matches the published results of \cite{Breit1928} for both excited and ground states. The maximum negative energy shift of order $\alpha^2/3$ is obtained for the ground state.

QFT corrections to the $g$-value of an electron are intrinsic and depend neither on the kinematics of the electron nor on the presence of an electromagnetic field,  while equation~\eqref{Hs} strongly supports the argument that relativistic corrections to the magnetic moment depend on both the kinematics of the electron and the electromagnetic field. The terms involving \(p^2\) and \(\nabla_B\) explicitly show that the correction depends on the motion of the electron and the spatial variation of the electromagnetic field. This dependence is universal and applies to both bound states (e.g., hydrogen-like atoms) and scattering systems, making equation~\eqref{Hs} a powerful tool for studying relativistic effects in various physical contexts such as Penning traps, muon storage rings, highly charged ions and scattering processes under strong electromagnetic fields.

Our results have direct implications for precision measurements in Penning traps, where the $\nabla_B$ terms in $H_s$ (equation~\eqref{Hs}) are particularly relevant for characterizing field misalignment effects in $g-2$ experiments. These terms could help quantify and reduce systematic uncertainties arising from imperfect field alignment, thereby improving the accuracy of $g$-factor measurements.

The $m_j^2$-dependent relativistic correction can be directly tested in Penning trap experiments through differential measurements of the $g$-factor for different $m_j$ states, as described in Section~\ref{subsec:penning}. For muonic hydrogen atoms, where relativistic effects are amplified by the muon's larger mass ($m_\mu \approx 207 m_e$), further investigation requires extending our Hamiltonian to include proton recoil effects, finite proton size corrections, and the spin-spin interaction between the muon and proton. Such an extension could provide new insights into the proton radius puzzle through precision measurements of the Lamb shift in muonic hydrogen systems, where the enhanced sensitivity to proton structure effects offers unique opportunities to test fundamental physics beyond the standard model \cite{Barger2011}.

\appendix

\section{Appendix A:Basic Definitions and Mathematical Identities}
\label{app:definitions}

This appendix collects fundamental definitions and identities used throughout the main text.

\subsection{Pauli Matrices and Identities}

The Pauli matrices are defined as:
\begin{align}
\sigma_1 &= \begin{pmatrix}0&1\\1&0\end{pmatrix}, &
\sigma_2 &= \begin{pmatrix}0&-i\\i&0\end{pmatrix}, &
\sigma_3 &= \begin{pmatrix}1&0\\0&-1\end{pmatrix}
\end{align}

The fundamental Pauli identity states:
\begin{equation}
(\vec{\sigma} \cdot \vec{a})(\vec{\sigma} \cdot \vec{b}) = \vec{a} \cdot \vec{b} + i\vec{\sigma} \cdot (\vec{a} \times \vec{b})
\end{equation}

\subsection{Gamma Matrices}

In the chiral (Weyl) representation used throughout this work, the gamma matrices are:
\begin{align}
\gamma^0 &= \begin{pmatrix}1&0\\0&-1\end{pmatrix}, &
\gamma^i &= \begin{pmatrix}0&\sigma^i\\-\sigma^i&0\end{pmatrix}
\end{align}

\subsection{Relativistic Energy-Momentum Relations}

The invariant mass $m$ is defined by the energy-momentum relation:
\begin{equation}\label{A_invariant_mass}
mc = \sqrt{p^\mu p_\mu} = \sqrt{\left(\frac{E}{c}\right)^2 - |\vec{p}|^2 c^2}.
\end{equation}

The expansion of relativistic energy with respect to rest mass energy is given by:
\begin{equation}\label{A_energy_expansion}
E = mc^2\sqrt{1 + \frac{p^2}{m^2c^2}} \approx mc^2 + \frac{p^2}{2m} - \frac{p^4}{8m^3c^2} + \cdots
\end{equation}

\subsection{Other Fundamental Relations}

The covariant derivative is defined as:
\begin{equation}\label{eq:covarient}
D_\mu(x) = \partial_\mu + \frac{ie}{\hbar c}A_\mu(x).
\end{equation}

The canonical momentum in electromagnetic fields:
\begin{equation}
\vec{\pi} = \vec{p} - \frac{e}{c}\vec{A}.
\end{equation}

The following anticommutation relation for arbitrary operators $A$ and $B$ is used in the main text:
\begin{equation}\label{eq:anticommutation}
\{A^2, B\} - 2ABA = [A, [A, B]]
\end{equation}

\section*{Appendix B: Calculation of $\frac{(\vec{\sigma} \cdot \vec{\pi})^4}{8m^3 c^2}$}

Start with
\begin{equation}\label{B-1}
(\vec{\sigma} \cdot \vec{\pi})^2 = \left(\vec{p} - \frac{e \vec{A}}{c}\right)^2 - \frac{e \hbar}{c} \vec{\sigma} \cdot \vec{B}. \tag{B-1}
\end{equation}
Here, we assume the momentum operator $\vec{p}$ acts on an implicit wavefunction, while the gradient operator $\vec{\nabla}$ with a subscript acts explicitly on external fields. To clarify this distinction, consider:

\begin{equation}\label{B-2}
\left(\vec{p} - \frac{e \vec{A}}{c}\right)^2 = \left(\vec{p} - \frac{e \vec{A}}{c}\right)\left(\vec{p} - \frac{e \vec{A}}{c}\right) = p^2 + i\hbar \vec{\nabla}_A \cdot \vec{A} - \frac{2e}{c} \vec{A} \cdot \vec{p} + \frac{e^2}{c^2} A^2. \tag{B-2}
\end{equation}
Under the Coulomb gauge condition $\vec{\nabla}_A \cdot \vec{A} = 0$, Eq.~\eqref{B-1} simplifies to:

\begin{equation}\label{B-3}
(\vec{\sigma} \cdot \vec{\pi})^2 = p^2 - \frac{2e}{c} \vec{A} \cdot \vec{p} + \frac{e^2}{c^2} A^2 - \frac{e \hbar}{c} \vec{\sigma} \cdot \vec{B}. \tag{B-3}
\end{equation}

Next, we apply $(\vec{\sigma} \cdot \vec{\pi})$ to both sides of Eq.~\eqref{B-3}, keeping only terms of order $1/c$:

\begin{align}\label{B-4}
(\vec{\sigma} \cdot \vec{\pi})^3 ={} & \vec{\sigma} \cdot \vec{p} \, p^2 + \frac{2i e \hbar}{c} (\vec{\sigma} \cdot \vec{\nabla}_A)(\vec{A} \cdot \vec{p}) - \frac{2e}{c} (\vec{A} \cdot \vec{p})(\vec{\sigma} \cdot \vec{p}) \nonumber \\
& -\frac{e\hbar}{c}(\vec{p}\cdot \vec{B} + \hbar \vec{\sigma}\cdot(\vec{\nabla}_B \times \vec{B}) + i\vec{\sigma}\cdot(\vec{p} \times \vec{B})) - \frac{e}{c} \vec{\sigma} \cdot \vec{A} p^2. \tag{B-4}
\end{align}

We have used the Pauli identity:
\begin{equation}\label{B-5}
\sigma_i \sigma_j = \delta_{ij} + i \epsilon_{ijk} \sigma_k, \tag{A-5}
\end{equation}
where $\epsilon_{ijk}$ is the Levi-Civita symbol. Since $\frac{(\vec{\sigma} \cdot \vec{\pi})^4}{8m^3 c^2}$ is suppressed by $1/c^2$, we drop higher-order terms beyond this precision in equation ~\ref{B-4}. Applying $\vec{\sigma} \cdot \vec{\pi}$ again and simplifying, we obtain:

\begin{align}\label{B-6}
\frac{(\vec{\sigma} \cdot \vec{\pi})^4}{8m^3c^2} ={} & \frac{p^4}{8m^3c^2} - \frac{4e}{8m^3c^3}\left(p^2 - i\hbar (\vec{p} \cdot \vec{\nabla}_{A}) - \frac{ \hbar^2}{2} \nabla_{A}^2 \right)\left(\vec{A} \cdot \vec{p}\right) \nonumber \\
& - \frac{2e\hbar}{8m^3c^3} \left(p^2 - i\hbar (\vec{p} \cdot \vec{\nabla}_B) - \frac{\hbar^2}{2} \nabla^2_B \right)(\vec{\sigma} \cdot \vec{B}). \tag{B-6}
\end{align}

The following identities are used in the derivation:

\begin{align*}
\vec{p} \times (\vec{B} \times \vec{p}) &= p^2 \vec{B} - (\vec{p} \cdot \vec{B}) \vec{p}, \\
\vec{\nabla} \times (\vec{f} \times \vec{g}) &= (\vec{g} \cdot \vec{\nabla}) \vec{f} - \vec{g} (\vec{\nabla} \cdot \vec{f}) - (\vec{f} \cdot \vec{\nabla}) \vec{g} + \vec{f} (\vec{\nabla} \cdot \vec{g}), \\
\vec{\nabla}_{B} \times (\vec{B} \times \vec{p}) &= (\vec{p} \cdot \vec{\nabla}_B) \vec{B}, \\
\vec{\nabla}_{B} \times (\vec{\nabla}_B \times \vec{B}) &= - \nabla^2_B \vec{B}, \\
\vec{\nabla}_{B} \cdot (\vec{B} \times \vec{p}) &= \vec{p} \cdot (\vec{\nabla}_B \times \vec{B}), \\
\vec{p} \times (\vec{\nabla}_B \times \vec{B}) &= \vec{\nabla}_B (\vec{p} \cdot \vec{B}) - (\vec{p} \cdot \vec{\nabla}_B) \vec{B}.
\end{align*}

\section*{Appendix C: Calculation of $\frac{[\vec{\sigma} \cdot \vec{\pi}, [\vec{\sigma} \cdot \vec{\pi}, (E_R - eA_0)]]}{8m^2 c^2}$}

The fourth term in equation ~(\ref{3.16}) can be evaluated by substituting $E_R = \frac{(\vec{\sigma} \cdot \vec{\pi})^2}{2m}$. The inner commutator is straightforward to compute:

\begin{align}\label{C-1}
[\vec{\sigma} \cdot \vec{\pi}, \frac{(\vec{\sigma} \cdot \vec{\pi})^2}{2m} - eA_0] &= -[\vec{\sigma} \cdot \vec{p}, eA_0] = i e \hbar \vec{\sigma} \cdot \vec{\nabla}_{A_0} A_0. \tag{B-1}
\end{align}

Now consider the outer commutator in detail:

\begin{align}\label{C-2}
[\vec{\sigma} \cdot (\vec{p} - \frac{e\vec{A}}{c}), i e \hbar \vec{\sigma} \cdot \vec{\nabla}_{A_0} A_0] &= e \hbar^2 \nabla^2_{A_0} A_0 + i e \hbar^2 \vec{\sigma} \cdot (\vec{\nabla}_{A_0} \times \vec{\nabla}_{A_0} A_0)\nonumber \\
&\quad + \frac{2 e^2 \hbar}{c} \vec{\sigma} \cdot (\vec{A} \times \vec{\nabla}_{A_0} A_0) - 2 i e \hbar \vec{\sigma} \cdot (\vec{p} \times \vec{\nabla}_{A_0} A_0). \tag{C-2}
\end{align}

We have used Eq.~\eqref{B-5} and the following electromagnetic relations:
\begin{itemize}
  \item $\vec{\nabla}_{A_0} A_0 = -\left( \vec{E} + \frac{\partial \vec{A}}{\partial t} \right)$,
  \item $\vec{\nabla} \times \vec{E} = -\frac{\partial \vec{B}}{\partial t}$,
  \item $\vec{\nabla} \times \vec{A} = \vec{B}$,
  \item Coulomb gauge: $\vec{\nabla} \cdot \vec{A} = 0$.
\end{itemize}

Substituting these into Eq.~\eqref{C-2}, we obtain:

\begin{align}\label{C-3}
[\vec{\sigma} \cdot (\vec{p} - \frac{e\vec{A}}{c}), i e \hbar \vec{\sigma} \cdot \vec{\nabla}_{A_0} A_0] &= - e \hbar^2 \vec{\nabla} \cdot \vec{E} - \frac{2 e^2 \hbar}{c} \vec{\sigma} \cdot (\vec{A} \times \vec{E}) - \frac{2 e^2 \hbar}{c} \vec{\sigma} \cdot \left( \vec{A} \times \frac{\partial \vec{A}}{\partial t} \right) \nonumber \\
&\quad + 2 e \hbar \vec{\sigma} \cdot (\vec{p} \times \vec{E}) + 2 e \hbar \vec{\sigma} \cdot \left( \vec{p} \times \frac{\partial \vec{A}}{\partial t} \right). \tag{B-3}
\end{align}

Finally, the full double commutator evaluates to:

\begin{align}\label{C-4}
\frac{[\vec{\sigma} \cdot \vec{\pi}, [\vec{\sigma} \cdot \vec{\pi}, (E_R - eA_0)]]}{8m^2 c^2} &= -\frac{e \hbar^2}{8m^2 c^2} \vec{\nabla} \cdot \vec{E} - \frac{2 e^2 \hbar}{8m^2 c^3} \vec{\sigma} \cdot (\vec{A} \times \vec{E}) - \frac{2 e^2 \hbar}{8m^2 c^3} \vec{\sigma} \cdot \left( \vec{A} \times \frac{\partial \vec{A}}{\partial t} \right) \nonumber \\
&\quad + \frac{2 e \hbar}{8m^2 c^2} \vec{\sigma} \cdot (\vec{p} \times \vec{E}) + \frac{2 e \hbar}{8m^2 c^2} \vec{\sigma} \cdot \left( \vec{p} \times \frac{\partial \vec{A}}{\partial t} \right). \tag{C-4}
\end{align}

\end{document}